\documentclass[twocolumn,aps,prx,showpacs,amsmath,amssymb,floatfix,longbibliography,superscriptaddress,eprint]{revtex4-1}
\usepackage{subfig}
\usepackage{textcomp}
\usepackage[justification=Justified,
   format=plain]{caption} 
\usepackage{subcaption}
\usepackage{xcolor}
\definecolor{darkblue}{RGB}{0,0,150}
\definecolor{nightblue}{RGB}{0,0,100}

\usepackage{graphicx,mathtools,bm,bbm}
\usepackage{MnSymbol}
\usepackage{varwidth}
\usepackage[
colorlinks,
citecolor=darkblue,
linkcolor=darkblue,
urlcolor=nightblue]{hyperref}
\usepackage{nicefrac,xfrac}
\usepackage{tikz-feynman}

\usepackage[english]{babel}
\usepackage[babel,kerning=true,spacing=true]{microtype}
\usepackage[utf8]{inputenc}

\usepackage{soul}
\usepackage{braket}

\usepackage{makecell}   
\usepackage{array}      
\usepackage[margin=1in]{geometry}

\providecommand{\tabularnewline}{\\}

\begin{document}

\title{Charge and pair density waves in a spin and valley-polarized system at a Van-Hove singularity}

\author{Avigail Gil}
\affiliation{Department of Condensed Matter Physics,
Weizmann Institute of Science,
Rehovot 76100, Israel}

\author{Erez Berg}
\affiliation{Department of Condensed Matter Physics,
Weizmann Institute of Science,
Rehovot 76100, Israel}

\begin{abstract}
We study a single component (i.e., single valley, spin-polarized)
two-dimensional electron gas with $C_{3v}$ symmetry tuned to a Van-Hove (VH) singularity.
Generically, there may be either three or six VH points at the Fermi level, related to each other by symmetry. 
Using a renormalization group analysis, we show that when the effective interactions between electrons at the VH points are positive, the system is stable. In contrast, if the effective interactions are negative, the system develops an instability toward either pair density wave (PDW) or charge density wave (CDW) orders, depending on the anisotropy of the dispersion at the VH points. 
The PDW may have either a single wavevector or multiple wavevectors. 
The PDW phase with three coexisting wavevectors can support fractional $\tfrac{h}{6e}$ vortices.  
The interplay between the geometry of the Fermi surface and the singularity of the density of states is the key that enables PDW formation.
\end{abstract}

\maketitle

\section{Introduction}
In the vast majority of superconductors, the center of mass momentum of the Cooper pairs is zero. This a consequence of the fact that, in the presence of time reversal symmetry, even a small zero-momentum superconducting order parameter is sufficient to gap the entire Fermi surface (with the possible exception of nodal points in two spatial dimensions). 
Superconductors whose Cooper pair condensate carries a non-zero momentum are known as pair density wave (PDW) states~\cite{PhysRevB.79.064515,agterberg2020physics}. 
Such exotic states have been predicted to occur 
when time reversal is broken by a Zeeman field, just before being completely destroyed. In this context, they are known as Fulde-Ferrel-Larkin-Ovchinikov (FFLO) states~\cite{Fulde1964,larkin1964nonuniform}. 
Evidence for PDW states in the absence of explicit time reversal breaking has been reported in certain cuprates~\cite{Li2007,Berg2007,Tranquada2008,edkins2019magnetic} and other materials~\cite{Li2023,liu2023pair,gu2023detection,aishwarya2023magnetic,Devarakonda2024}. 

Generically, a PDW cannot emerge when the interaction strength is weak. Hence, PDWs are usually challenging to obtain theoretically, since they cannot be studied by perturbative techniques. 
Moreover, in this regime, the PDW state often competes with other forms of electronic order, such as charge density wave (CDW) states. Nevertheless, PDW order has been found in certain models of interacting electrons~\cite{berg2009striped,Berg2010,PhysRevLett.114.197001,Vafek2014,venderley2019evidence,peng2021precursor,Han2022,Zhang2022,huang2022pair,Schaffer2023,Wu2023,Jiang2024,PhysRevB.107.045122}.  

It thus came as a surprise when superconductivity was recently discovered in rhombohedral four and five-layer graphene (R4G and R5G) in a regime where the spin and valley symmetries are spontaneously broken in the normal state~\cite{han2024signatures}. The most straightforward interpretation of the experiment suggested that the superconducting state is fully spin and valley polarized. Such a state is most likely a PDW, since only states in a single valley are available for pairing near the Fermi energy. Several theoretical works appeared attempting to explain the emergence of this state, predicting a PDW state whose wavevector is either commensurate~\cite{chou2024DasSarmaTheory,geier2024chiraltopologicalsuperconductivityisospin,parramartinez2025bandrenormalizationquartermetals,Kim2025,shavit2024} or incommensurate~\cite{Yang2024,chen2025intrinsicsuperconductingdiodeeffect,gaggioli2025} with the underlying graphene lattice.

In this work, we investigate the types of electronic orders that develop when a fully spin and valley polarized two-dimensional system with $C_{3v}$ symmetry is tuned to the vicinity of a Van-Hove (VH) singularity. 
We study two generic situations with either three or six distinct VH points at the Fermi energy, related to each other by $C_{3v}$.
At the VH singularity, the density of states diverges logarithmically, allowing us to access the electronic instabilities in the weak coupling limit. Moreover, unlike in previous studies of spinful two-dimensional systems near a VH singularity~\cite{Dzyaloshinskii,H.J.Schulz_1987,Lederer,PhysRevLett.81.3195,nandkishore2012chiral,PhysRevB.98.205151,
PhysRevB.102.085103,PhysRevB.102.085103,Wu2023}, our spinless system exhibits only logarithmic divergences in both the particle-particle and particle-hole bare susceptibilities (in the spinful case, the susceptibility in the particle-particle channel typically diverges as the square of the logarithm of the low-energy cutoff~\cite{Dzyaloshinskii,H.J.Schulz_1987,Lederer}). Thus, in the spinless case, the two channels can be treated on equal footing, allowing for a better controlled solution of the problem.

Using a weak-coupling renormalization group (RG) analysis, we  find that weak repulsive effective interactions at the VH points are marginally irrelevant, and hence the system is stable even when tuned to the VH singularity. For weak attractive interactions, the dominant instability in the low energy limit is determined by the mass anistropy at the VH points and, in the case of six VH points, also on the angle between the principal axes between neighboring mirror symmetry related VH points. For small anistropy, the the CDW phase is favored. In contrast, for large anistropy, the system presents an instability towards PDW order. We discuss possible implications of our results for superconductivity and competing orders in rhombohedral multilayer graphene.  

\section{Three VH points}
\subsection{The Model}
\begin{figure}[ht]
    \centering
    \includegraphics[width= \columnwidth]{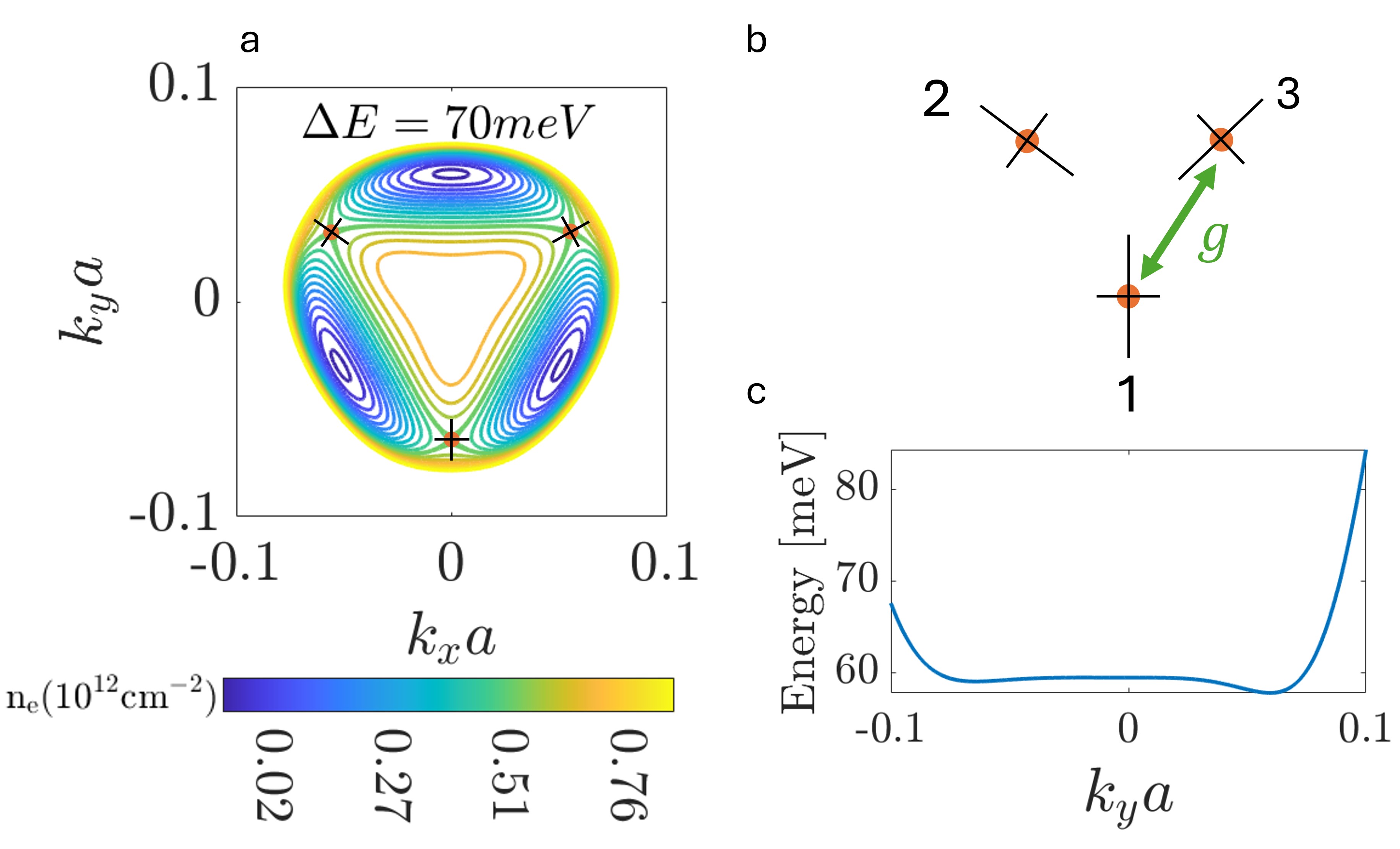}
    \caption{(a) \textbf{Iso-energetic lines for R4G} in the +K valley conduction band with a potential difference of $\Delta E = \text{70meV}$ between the top and bottom layers. The model parameters are taken from~\cite{chou2024DasSarmaTheory}. 
    Here, $a\approx 0.246$nm is the lattice constant of graphene. 
    The three distinct VH points are drawn schematically in orange on the contours and the black lines represent the principal axes of the saddle points. The VH points are reached for an density of $\rm{n_e}\approx 0.5\cdot 10^{12}$cm$^{-2}$. (b) \textbf{Interactions between electrons near the three VH points.} (c) \textbf{Energy cut vs. $k_y$ for $k_x=0$ of the conduction band of R4G.}} \label{fig: the model}
\end{figure}

 We study a $C_{3v}$ symmetric and spin and valley polarized model with three symmetry-related VH points at the Fermi energy. 
 We are interested in the weak-coupling instabilities that occur in the low energy limit. Since the density of states diverges logarthimically near the three VH points, we can treat a simpler problem that includes patches near these points.
 This type of behavior is exhibited by tuning the density and the displacement field in systems such as twisted bilayer graphene (TBG)~\cite{PhysRevB.98.205151}, twisted double bilayer graphene (TDBG)~\cite{PhysRevB.102.085103}, rhomobohedral trilayer graphene (R3G)~\cite{Zhou_2021, Chatterjee2022} and R4G~\cite{han2024signatures}. As an illustration, we show the band structure of R4G in Fig.~\ref{fig: the model}a. As can be seen in the figure, at a certain density that depends on the perpendicular displacement field, there are three symmetry-related VH points at the Fermi energy. 
 
 We thus consider a model that includes three patches centered at the three VH points at momenta $\vec{q}_\alpha$, $\alpha=1,2,3$, with one coupling constant $g$ between two electrons in any two different patches, as shown in Fig.~\ref{fig: the model}b.
In this model, there is no momentum-independent intra-patch interaction due to the full spin polarization.
There is a single inter-patch interaction parameter, $g$, as a result of the symmetry of the system. 
Intra and inter-patch momentum dependent interactions are 
irrelevant under RG, and will be neglected here. 
We will consider both signs of the coupling $g$. 
In Appendix~\ref{App E}, we argue that even if the bare interaction is purely repulsive (e.g., Coulomb interactions), the effective low-energy interaction $g$ may be negative. 

The low energy Hamiltonian in the patch approximation in the three VH points model is given by:
    \begin{align} 
    H_3 &= \int d^2 x \left[\sum_{\alpha=1}^3\psi_\alpha^\dag(\varepsilon_\alpha(\vec{k}) - \mu)\psi_\alpha
        +g\sum_{\alpha<\beta} \psi_\alpha^\dag \psi_{\beta}^\dag \psi_{\beta} \psi_\alpha \right], \label{eq: 1}
    \end{align} 
where $\psi_\alpha (\vec{x})$ is the fermionic field for an electron on patch $\alpha=1,\ 2,\ 3$. The low energy dispersion around the VH point labeled by $\alpha$ is described by $\varepsilon_\alpha(\vec{k})$, where $\vec{k} = -i\vec{\nabla}$ and $\mu$ is the chemical potential. 
In our model, we assume the mass at the VH point is anisotropic and is characterized by a mass anisotropy parameter, $\eta$. 
Defining 
\begin{align}
\label{eq: 2}
    \varepsilon_{\vec{k}} (\eta,\varphi) = &-\frac{\sqrt{\eta}}{2m} \left(\cos(\varphi)k_x -\sin(\varphi)k_y\right)^2 \\
    &+\frac{1}{2m\sqrt{\eta}}
    \left(\sin(\varphi)k_x +\cos(\varphi)k_y\right)^2 , \notag
\end{align}
the low energy dispersions near the three VH points at $\vec{q}_\alpha$, labeled by $\alpha=1,2,3$, are given by 
\begin{equation}
    \label{eq: 3}
    \varepsilon_{\alpha}\left(\vec{k}\right) = \varepsilon_{\vec{k} - \vec{q}_\alpha}\left(\eta,\varphi_\alpha=-\frac{2\pi}{3}(\alpha-2)\right).
\end{equation}
Since the behavior of the system for $\eta$ is identical to the behavior for $\frac{1}{\eta}$, we restrict our attention to $0<\eta<1$.  
We note that for $\eta = \frac{1}{3}$, we recover the structure of honeycomb tight binding model tuned to the VH singularity~\cite{nandkishore2012chiral}. 

\subsection{RG Analysis} \label{Section 1B}
We perform a renormalization group analysis of the Hamiltonian \eqref{eq: 1} by writing the equivalent action and imposing a cutoff in frequency. I.e. 
$\vec{k}$ is unbounded, but $\omega$ is bounded between $-\Lambda_\omega$ to $\Lambda_\omega$. 
In each RG step we integrate out modes with frequency $\Lambda_{\omega}/s<\omega<\Lambda$ and all momentum, and then rescale the frequency to recover the same cutoff as before.  
The momenta and frequency scale as:
\begin{subequations}  \label{eq: 4}
    \begin{align}
        &\omega' = s\omega,\\
        &k' = s^{\nicefrac{1}{2}}k,\\
        &\psi'(k',\theta,\omega') = s^{-2}\psi(k,\theta,\omega),
    \end{align}
\end{subequations}
where $\vec{k} = k (\cos\theta, \sin \theta)$. 

\begin{figure}[ht]
    \centering
    \includegraphics[width= \columnwidth]{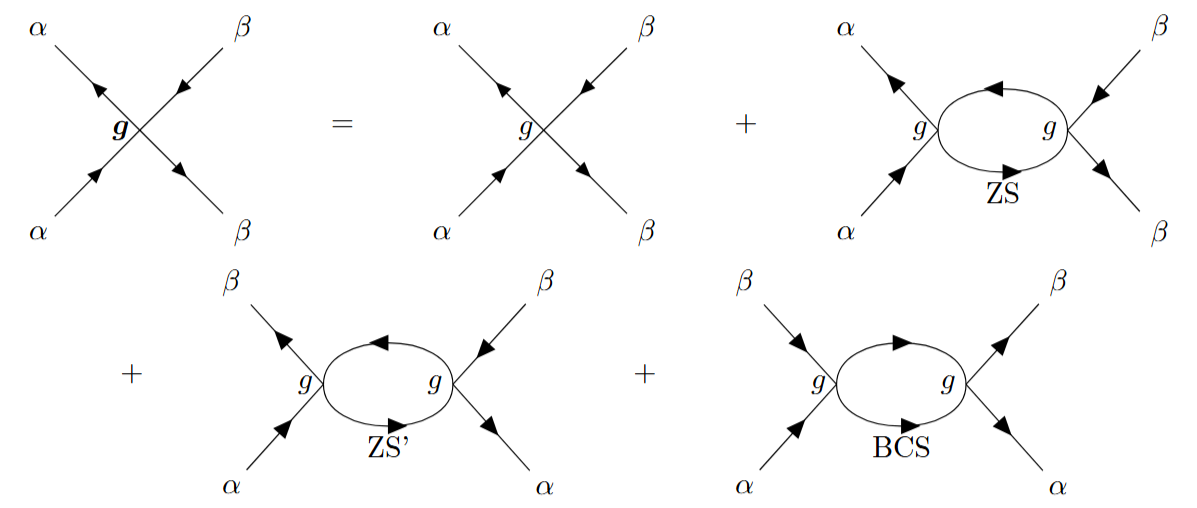} 
    \caption{\textbf{Diagrams contributing to the renormalization of $\boldsymbol{g}$} up to one loop order. The indices $\alpha$ and $\beta$ label the different patches (VH points). The outer frequencies are set to zero, and the frequency inside the loops lies in a shell of width $d\Lambda_\omega$ being integrated over. }
    \label{fig:int diags}
\end{figure}

At the tree level, the integration out of the modes and the rescaling of the field leaves the action invariant.
To one-loop order, there are three contributions to the renormalization of the coupling constant, known as the BCS diagram, the ZS diagram and the ZS' diagram~\cite{RevModPhys.66.129}, shown in Fig.~\ref{fig:int diags}.
We find that the ZS diagram does not contribute to the $\beta$ function since both of the poles are on the same half plane.
The remaining ZS' diagram and BCS diagram each contribute to the flow.
The details of the calculation can be found in Appendix~\ref{App A}. To one-loop order, the RG flow for the interaction $g$ is given by
\begin{equation}
    \Dot{g} = -a(\eta)g^2 \equiv -\left(a_{\text{ZS'}}(\eta) + a_{\text{BCS}}(\eta)\right)g^2  \label{eq: 5}
\end{equation}
where the flow parameter is $dt = d\log\left(\Lambda_\omega\right)$. The full expressions for $a_{\text{ZS'}}(\eta)$ and $a_{\text{BCS}}(\eta)$ are also given in Appendix~\ref{App A}.

For our purpose, the important characteristics of $a(\eta)$ are that it is positive for all $\eta$ and it diverges at $\eta=\tfrac{1}{3}$. For $\eta = \tfrac{1}{3}$, the contribution of the particle-hole channel has a log-squared divergence, due to the perfect nesting of the Fermi surface. 
Thus, near $\eta = \tfrac{1}{3}$ our description breaks down. 
An analysis of the case of $\eta = \frac{1}{3}$ for spinful fermions can be found in~\cite{PhysRevB.102.085103}.

Solving Eq. \eqref{eq: 5}, We find that for an initial repulsive interaction ($g>0$), $g$ flows towards zero and the system is stable. For an initial attractive interaction, $g < 0$, the interaction diverges under the RG flow. $g(t)$ is given by:
\begin{equation}
    g(t) = \frac{1}{a(\eta)\left(t-t_c\right)}, \label{eq: 6}
\end{equation}
 where $t_c = \tfrac{1}{a(\eta)|g(0)|}$, and $E_c = \Lambda_0 \exp(-1/t_c)$ is the critical energy scale where the perturbative treatment breaks down (here, $\Lambda_0$ is the initial upper cutoff of the theory).

 \subsection{Instabilities}

We study the various instabilities the model develops at the energy scale $E_c$. To probe the tendency to develop each order, we add the various test vertices to the Hamiltonian and track their RG flow. 

In the three VH point model, the possible test vertices are
\begin{equation} \label{eq: 7}
    \rho_{\alpha}^{(0)}\psi_\alpha^\dagger\psi_\alpha,  \quad \rho_{\alpha,\beta}\psi_\alpha^\dagger\psi_\beta
    \,(\alpha\ne\beta),  \quad \Delta_{\alpha,\beta}\psi_\alpha^\dagger\psi_\beta^\dagger.
\end{equation}
The diagrams that contribute to their renormalization are shown in Appendix~\ref{App B}. 
The test vertices in Eq. \eqref{eq: 7} correspond to different types of electronic order: $\rho_{\alpha}^{(0)}$ is a uniform (zero-wavevector) charge order, $\rho_{\alpha,\beta}$ is a charge density wave between patches $\alpha$ and $\beta$ with wavevector $\vec{q}_\alpha - \vec{q}_\beta$, and $\Delta_{\alpha,\beta}$ is a pair density wave between $\alpha$ and $\beta$ whose wavevector is $\vec{q}_\alpha + \vec{q}_\beta$. We find that the test vertices for each instability renormalize with RG in the following way to one loop order:
\begin{subequations}
\label{eq: 8}
    \begin{align}
        &\frac{d\rho_{\alpha}^{(0)}}{dt} = 0,\\
        &\frac{d\rho_{\alpha,\beta}}{dt} = -g(t)a_\text{ZS'}(\eta)\rho_{\alpha,\beta},\\
        &\frac{d\Delta_{\alpha,\beta}}{dt} = -2g(t)a_\text{BCS}(\eta)\Delta_{\alpha,\beta}.
    \end{align}
\end{subequations}

From Eqs. \eqref{eq: 8}, we find that uniform charge order is marginal to one loop, while the CDW and PDW orders are suppressed or enhanced depending on the sign of $a_\text{ZS'}(\eta)$ and $a_\text{BCS}(\eta)$. 
\begin{figure}[hb]
    \centering
    \includegraphics[width= \columnwidth]{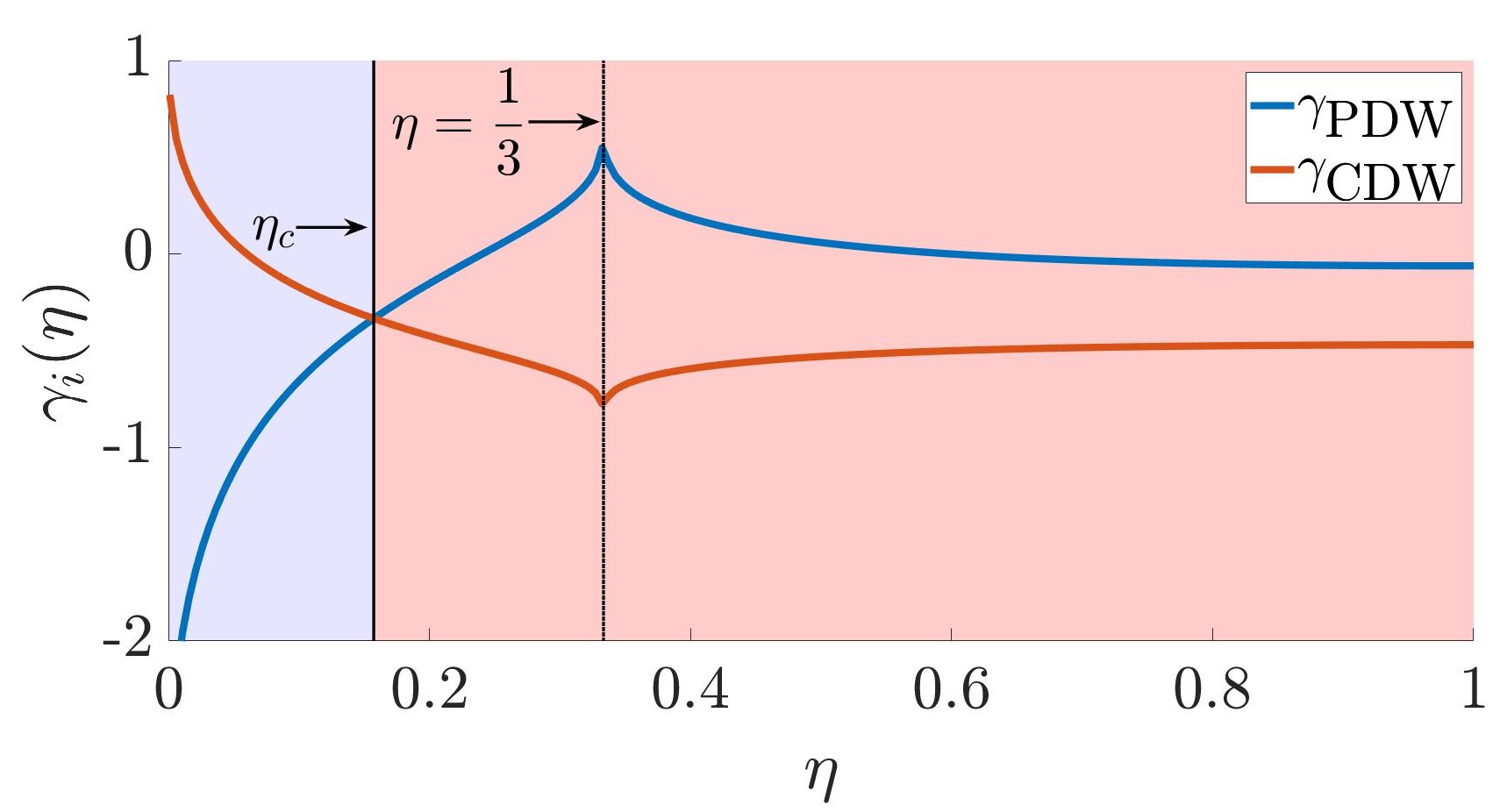} 
    \caption{\textbf{The exponents of the PDW and CDW susceptibilities}, $\chi_i\approx\left(t-t_c\right)^{\gamma_i}$. The inter-patch PDW is shown in blue and the CDW is drawn in orange. For $\eta<\eta_c$, the dominant instability is toward an inter-patch PDW order and for $\eta>\eta_c$ the CDW instability is dominant.}
    \label{fig: 3VHS SC sus}
\end{figure}
Solving for the flow of the vertices, assuming that close to $t_c$, the order parameter evolves as $\left(t-t_c\right)^{\alpha_i}$ ($i={\mathrm{CDW},\mathrm{PDW}}$), we get:
\begin{subequations}
\begin{align}
    &\alpha_\text{CDW}(\eta)=-\frac{a_\text{ZS'}(\eta)}{a(\eta)},\\
    &\alpha_\text{PDW}(\eta)=-2\frac{a_\text{BCS}(\eta)}{a(\eta)}.
\end{align} \label{eq: 9}
\end{subequations}
Near $t_c$, the susceptibilities diverge as $\chi_i\approx\left(t-t_c\right)^{\gamma_i}$ with $\gamma_i=2\alpha_i+1$ ~\cite{PhysRevB.100.085136}.
The exponents of the susceptibilities $\gamma_i(\eta)$ are plotted in Fig.~\ref{fig: 3VHS SC sus}. We find a crossover at $\eta_c\approx0.157$ where for $\eta < \eta_c$, the instability toward a PDW phase is more dominant.

\subsection{Competition between orders}

For $\eta < \eta_c$ ($\eta > \eta_c$) the system can either develop a single $q$ component PDW (CDW) phase or a multiple $q$ component PDW (CDW) phase, respectively. The system will tend toward the state that minimizes the free energy. 
We treat $\rho_{\alpha,\beta}$ and $\Delta_{\alpha,\beta}$ as order parameters and write the Ginzburg-Landau free energy functional near $T_c$, up to fourth order in the order parameters,
\begin{widetext}
    \begin{align} \label{eq: 10}
    F(&\Delta_{\alpha,\beta},\rho_{\alpha,\beta}) = \frac{1}{2}  \sum_{\alpha,\beta} a^\text{P}_{\alpha,\beta}(T-T_\text{P}) |\Delta_{\alpha,\beta}|^2 + \frac{1}{4} \sum_{\alpha,\beta \ \neq \ \gamma,\delta} b^\text{P}_{\alpha,\beta,\gamma,\delta} |\Delta_{\alpha,\beta}|^2 |\Delta_{\gamma,\delta}|^2 +
    \frac{1}{2} \sum_{\alpha,\beta} c^\text{P}_{\alpha,\beta} |\Delta_{\alpha,\beta}|^4 +\notag \\    
    &\frac{1}{4}\sum_{\alpha \neq \beta \neq \gamma} d^\text{P}_{\alpha,\beta,\gamma} \left(\Delta_{\alpha,\beta}\right)^* \Delta_{\gamma,\alpha}\rho_{\beta,\gamma}  + c.c. 
    + \frac{1}{4}\sum_{\alpha,\beta\,\gamma,\delta} u_{\alpha,\beta,\gamma,\delta} |\Delta_{\alpha,\beta}|^2|\rho_{\gamma,\delta}|^2 + \sum_{\alpha,\beta\ \neq\ \gamma,\delta}v_{\alpha,\beta, \gamma,\delta}\Delta_{\alpha,\beta}^*\Delta_{\gamma,\delta} \rho^*_{\gamma, \beta}\rho_{\alpha,\delta} +\\
    &\frac{1}{2} \sum_{\alpha,\beta} a^\text{C}_{\alpha,\beta}(T-T_\text{C}) |\rho_{\alpha,\beta}|^2 +
    \frac{1}{4} \sum_{\alpha,\beta\ \neq\ \gamma,\delta} b^\text{C}_{\alpha,\beta,\gamma,\delta} |\rho_{\alpha,\beta}|^2|\rho_{\gamma,\delta}|^2 +\frac{1}{2} \sum_{\alpha,\beta} c^\text{C}_{\alpha,\beta} |\rho_{\alpha,\beta}|^4 + \frac{1}{2} \sum_{\alpha \neq \beta \neq \gamma} d^\text{C}_{\alpha,\beta,\gamma} \rho_{\alpha,\beta}\rho_{\gamma,\alpha}\rho_{\beta,\gamma}+ c.c. \notag
    \end{align}
\end{widetext}
The calculation of the free energy coefficients $b_{\alpha,\beta,\gamma,\delta}^i$, $c^i_{\alpha,\beta}$, $d^i_{\alpha,\beta,\gamma}$, where $i=\text{P or C}$, and $u_{\alpha,\beta,\gamma, \delta}$, and $v_{\alpha,\beta,\gamma,\delta}$ from the microscopic Hamiltonian is given in Appendix~\ref{App C}. 
We note that $\rho_{\alpha,\beta} = \rho^*_{\beta,\alpha}$ and that $\Delta_{\alpha,\beta} = -\Delta_{\beta,\alpha}$.

We find that, for $\eta$ near $\eta_c$, the Ginzburg-Landau free energy is unbounded from below in the fourth order approximation (see Appendix~\ref{App C}). 
Analyzing the phase diagram thus requires including higher-order terms. 
We still note, however, that when the system is in a CDW state, it will tend to develop all three  components together (which we denote as a $3q$ CDW state).
The presence of the cubic term $d^\text{C}$ in the free energy, dictates that the transition to this state is first order. 
Since $d^\text{C}>0$, the free energy is minimized when 
the sum of the three phases is equal to $2\pi (n+\frac{1}{2})$.
This constraint leaves two independent phases, that transform under the generators of the two spontaneously broken translation symmetries. 

\begin{figure}[h!]
    \centering
    \includegraphics[width= \columnwidth]{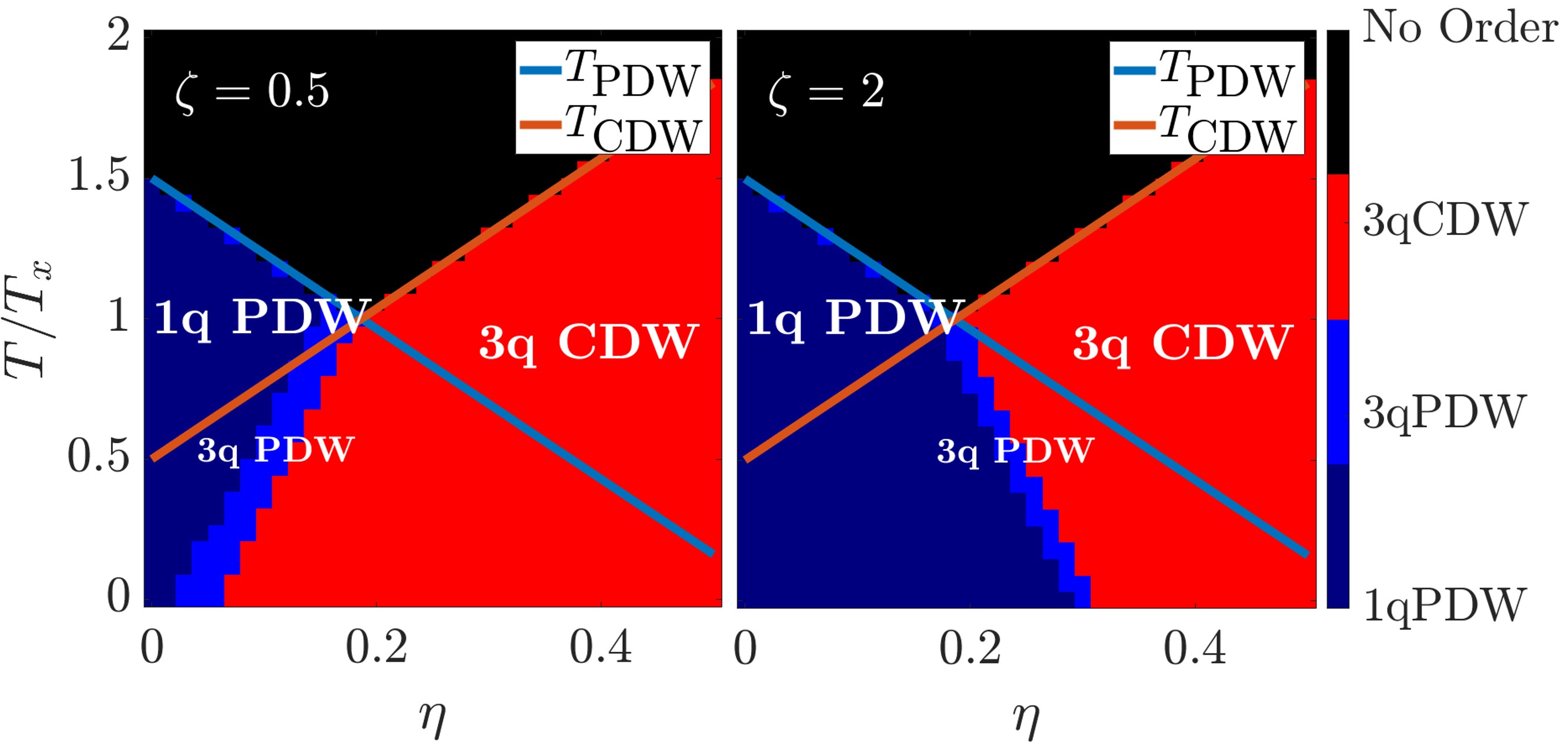} 
    \caption{\textbf{Qualitative phase diagrams for the three VH point model.} The GL free energy coefficients were set to their values at $\eta = \eta_c$. The coefficient of the phenomenological sixth-order term was set to $w=0.1$. $T_x$ is the temperature where $T_{\rm{PDW}} = T_{\rm{CDW}}$ cross at $\eta=\eta_c$. 
    We have assumed the form: $T_{\rm{PDW}}(\eta) = T_x[1 - \tfrac{1}{2\eta_c}(\eta - \eta_c)]$, $T_{\rm{CDW}}(\eta) = T_x[1 + \tfrac{1}{2\eta_c}(\eta - \eta_c)]$. 
    We note that as long as $w>0$, the phase diagram does not depend on its value. The boundary between the 3$q$ CDW phase to the 1$q$ PDW phase depends on the value of the phenomenological parameter $\zeta$.} 
    \label{fig: 3VHS bounded phase}
\end{figure}

Phenomenologically adding a sixth order term $w\left[\sum_{\alpha,\beta}(|\Delta_{\alpha,\beta}|^2+\zeta|\rho_{\alpha,\beta}|^2)\right]^3$ with $\zeta>0$ to the expression of the free energy given in Eq. \eqref{eq: 10} bounds it from below. Minimizing the free energy numerically with respect to the order parameters, we can obtain a qualitative phase diagram (see Fig.~\ref{fig: 3VHS bounded phase}).
We note that all of the phase transitions in this model are first order except the transition from the disordered to 1$q$ PDW which is second order.

\section{Six VH points}
\subsection{Model}

\begin{figure}[ht]
    \centering
    \includegraphics[width= \columnwidth]{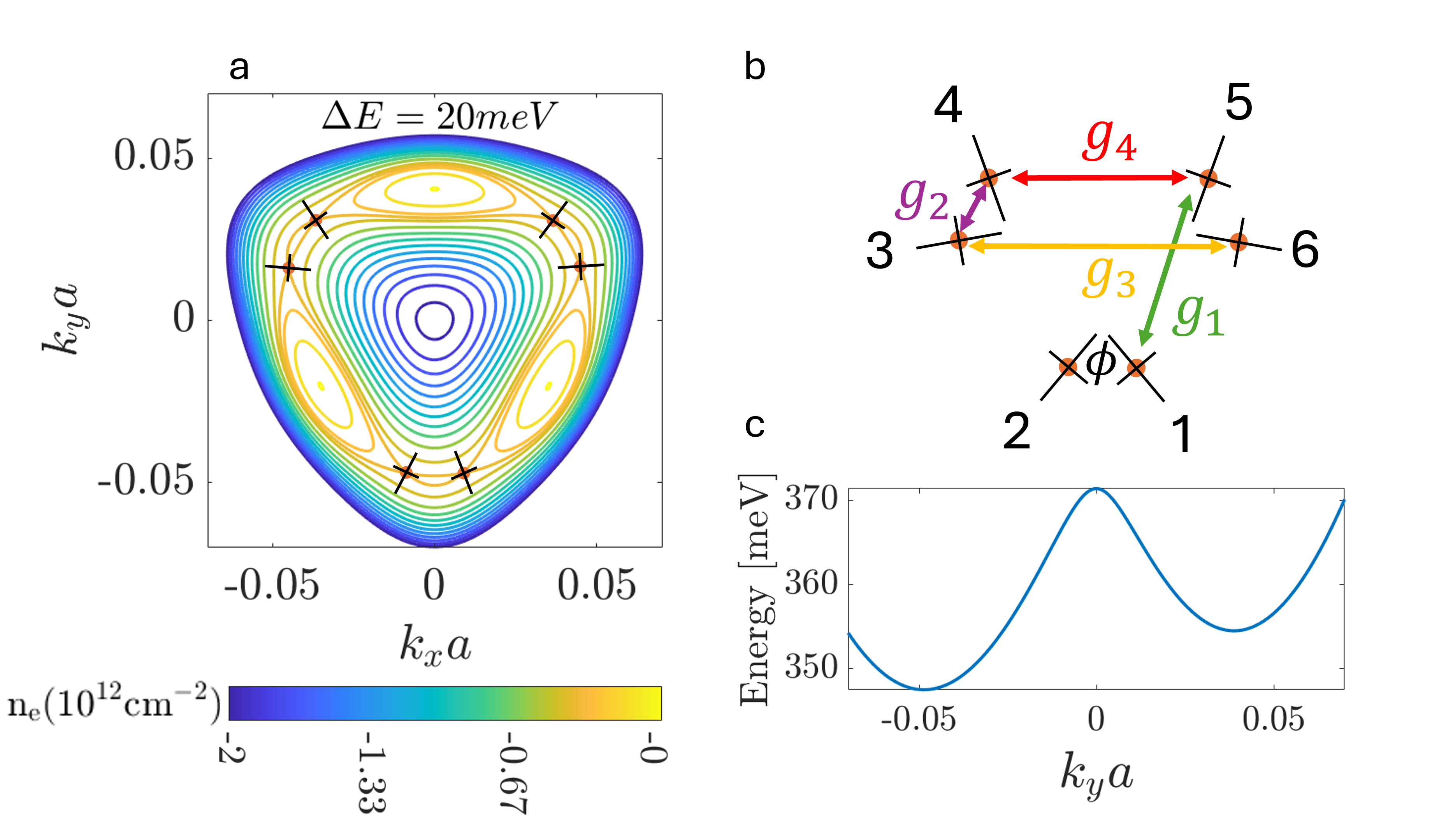}
    \caption{(a) \textbf{Iso-energetic lines for R3G vs. $\vec{k}$} in the +K valley conduction band with a potential difference $\Delta E = 20$meV between the top and bottom layers. The model parameters were taken from Ref.~\cite{Chatterjee2022} . The six distinct VH points are drawn schematically in orange on the contours and the black lines represent the principal axes of the saddle points. (b) \textbf{Schematics of the distinct inter-patch interactions in a spin polarized model with six VH points.} (c) \textbf{Energy cut vs. $k_y$ for $k_x=0$ of the conduction band of R3G.}} \label{fig: the second model}
\end{figure}

We now turn to analyze the second model with six VH points. This structure can arise by tuning the density and the displacement field in systems such as TBG~\cite{PhysRevB.98.205151}, TDBG~\cite{PhysRevB.102.085103} or R3G~\cite{Zhou_2021, Chatterjee2022}. An example of the band structure of R3G with six VH points is plotted in Fig.~\ref{fig: the second model}a. The VH points are related to each other either by $C_3$ rotational symmetry or by mirror symmetry. 

The low energy Hamiltonian in the patch approximation is given by:
\begin{equation} \label{eq: 11}
        H_6 = \int d^2 x \left[ \sum_{\alpha=1}^6 \psi_\alpha^\dag (\varepsilon_\alpha(\vec{k}) - \mu) \psi_\alpha +
        \sum_{\alpha<\beta} g_{\alpha,\beta} \psi_\alpha^\dag \psi_{\beta}^\dag \psi_{\beta} \psi_\alpha \right],
\end{equation}    
where $\psi_\alpha (\vec{x})$ is the fermionic field for an electron on patch $\alpha$. The interactions $g_{\alpha,\beta}$ are related to each other through the $C_{3v}$ symmetry of the model:
\begin{subequations}
    \begin{align}
    &g_{\alpha,\beta} = g_{\beta,\alpha},\\
    &g_{1,3}=g_{2,4}=g_{3,5}=g_{4,6}=g_{5,1}=g_{6,2}\equiv g_1,\\
    &g_{1,2}=g_{3,4}=g_{5,6}\equiv g_2,\\
    &g_{1,4}=g_{3,6}=g_{5,2}\equiv g_3,\\
    &g_{2,3}=g_{4,5}=g_{6,1}\equiv g_4.
    \end{align}
\end{subequations}
The low energy dispersion around the VH point labeled by $\alpha$ is described by $\varepsilon_\alpha(\vec{k})$ and $\mu$ is the chemical potential. 
The low energy dispersion around the six VH points is given by 
\begin{subequations}
    \label{eq: 12}
    \begin{align}
    \varepsilon_{2\alpha -1}\left(\vec{k}\right) &= \varepsilon_{\vec{k}-\vec{q}_\alpha} \left(\eta,\varphi_{2\alpha-1} =-\frac{2\pi}{3}(\alpha-2)\right),\\
    \varepsilon_{2\alpha}\left(\vec{k}\right) &= \varepsilon_{\vec{k}-\vec{q}_\alpha} \left(\eta,\varphi_{2\alpha}=-\frac{2\pi}{3}(\alpha-2)-\phi\right),
    \end{align} 
\end{subequations}
where $\varepsilon_{\vec{k}}(\eta,\varphi)$ is given in Eq. \eqref{eq: 2}.
Here, we introduced another parameter $0<\phi<\pi$, which is the angle 
between the principal axes of patches 1 and 2. 
(see Fig. \ref{fig: the second model}b). Due to the quadratic approximation of the energy dispersion, the parameter $\phi$ has a period of $\pi$. We note that for $\eta = \frac{1}{3}$ and $\phi = \frac{\pi}{3}$, we recover the structure of the VH points obtained in a  tight binding model on a honeycomb lattice. 

\subsection{RG Analysis}

Just as in the previous model, we turn to RG in order to analyze the stability of the model. The scalings of the fields and momenta are given in Eqs. \eqref{eq: 4}. To tree level, the integration out of the modes and the rescaling of the field leaves the action invariant.

Due to the fact that each interaction involves a momentum transfer of different magnitude, there is no mixing between the different couplings under the RG flow and the RG equations are decoupled.
Following a  calculation similar to the three VH case, the resulting RG equations can be written as:
\begin{equation} \label{eq: 13}
    \Dot{g}_{\alpha,\beta} = -A\left(\eta,\varphi_\alpha-\varphi_\beta\right)g_{\alpha,\beta}^2,
\end{equation}
where the coefficient $A\left(\eta,\varphi_\alpha-\varphi_\beta\right)$ depends on the anisotropy of the dispersion and the angle $\varphi_\alpha - \varphi_\beta$ between the principal axes of the two patches that participate in the coupling $g_{\alpha,\beta}$. 
As in the three VH case, we can write $A(\eta,\phi)$ as a sum of ZS' and BCS contributions:
\begin{equation}
    \label{eq: 15}
    A(\eta,\phi)=A^\text{ZS'}(\eta,\phi) + A^\text{BCS}(\eta,\phi).
\end{equation}
The full expressions for $A^\text{ZS'}(\eta,\phi)$ and $A^\text{BCS}(\eta,\phi)$ are given in Appendix~\ref{App D}. 

The flow coefficient $A(\eta,\phi)$ is positive for all $\eta$ and $\phi$, meaning that for initial repulsive interactions ($g_{\alpha,\beta}>0$), the system remains stable despite the diverging density of states. For initial attractive interactions, all of the interactions diverge under the RG flow for all $\eta$ and $\phi$ and the system becomes unstable.
We note that, analogously to the three VH case, $A(\eta,\phi=\frac{\pi}{3})$ has a divergence at $\eta=\frac{1}{3}$, due to the presence of a susceptibility that diverges as log squared of the IR cutoff.

\subsection{Instabilities}
To probe the tendency to develop each order, we add test vertices to the Hamiltonian and track how they flow. 
The test vertices and the diagrams that contribute to the renormalization in this model are similar to those in the previous three VH point model and can be found in Appendix~\ref{App B}. In this model, we need to consider the following test vertices:
\begin{equation} \label{eq: 16}
 \rho_{\alpha}^{(0)}\psi_\alpha^\dagger\psi_\alpha,  \quad \rho_{\alpha,\beta}\psi_\alpha^\dagger\psi_\beta,  \quad \Delta_{\alpha,\beta}\psi_\alpha^\dagger\psi_\beta^\dagger.
\end{equation}

We find that the vertices for each instability renormalize, up to one loop order, in the following way:
\begin{align}
    &\frac{d\rho_{\alpha}^{(0)}}{dt} = 0,\\
    &\frac{d\rho_{\alpha,\beta}}{dt} = -g_{\alpha,\beta}(t)A^\text{ZS'}(\eta,\varphi_\alpha -\varphi_\beta)\rho_{\alpha,\beta},\\
    &\frac{d\Delta_{\alpha,\beta}}{dt} = -2g_{\alpha,\beta}(t)A^\text{BCS}(\eta,\varphi_\alpha -\varphi_\beta)\Delta_{\alpha,\beta}.
\end{align}
The uniform charge order is marginal to one loop. The CDW and PDW phases are suppressed or enhanced depending on the sign of $A^\text{ZS'}(\eta,\varphi_\alpha -\varphi_\beta)$ and $A^\text{BCS}(\eta,\varphi_\alpha -\varphi_\beta)$. 
Solving for the flow of the vertices, assuming that close to the instability, $\Delta_{\alpha,\beta}$ evolves as $\left(t-t_{c;\alpha,\beta}\right)^{\alpha_{\alpha,\beta}}$, we get:
\begin{align}
    &\alpha_{\alpha,\beta}^\text{PDW}(\eta,\varphi_\alpha -\varphi_\beta)=-2\frac{A^\text{BCS}(\eta,\varphi_\alpha -\varphi_\beta)}{A(\eta,\varphi_\alpha -\varphi_\beta)}\\
    &\alpha_{\alpha,\beta}^\text{CDW}(\eta,\varphi_\alpha -\varphi_\beta)=-\frac{A^\text{ZS'}(\eta,\varphi_\alpha -\varphi_\beta)}{A(\eta,\varphi_\alpha -\varphi_\beta)}
\end{align}
Near $t_{c;\alpha,\beta}$, the susceptibilities evolve as $\chi_{\alpha,\beta}\approx\left(t-t_{c;\alpha,\beta}\right)^{\gamma_{\alpha,\beta}}$ with $\gamma_{\alpha,\beta}=2\alpha_{\alpha,\beta}+1$.

\begin{figure}[h!]
    \centering
    \includegraphics[width= \columnwidth]{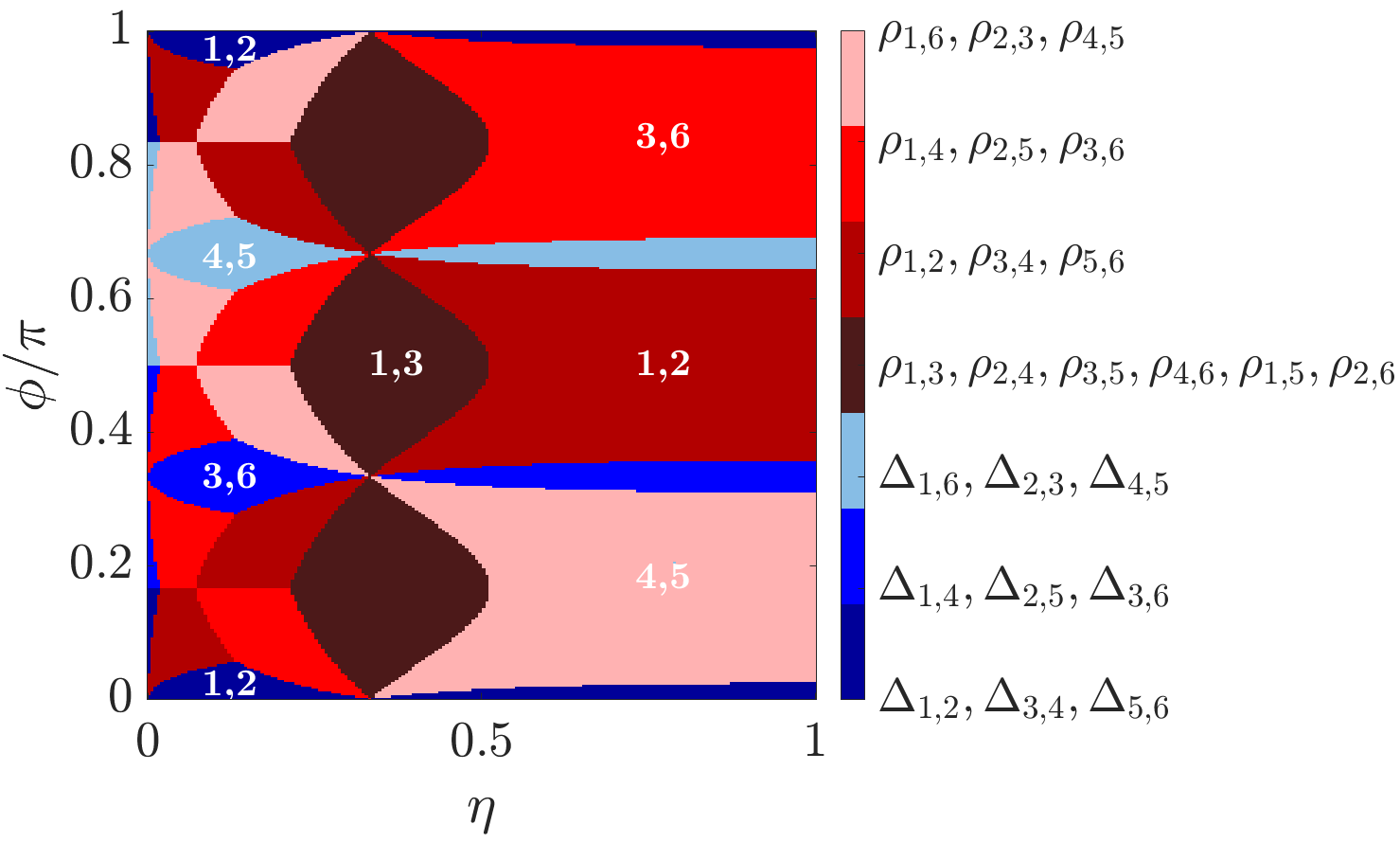} 
    \caption{\textbf{Leading instability of the six VH point model as a function of $\eta$ and $\phi$.}
     The type of order with the most divergent susceptibility is indicated for each $(\eta,\phi)$. 
     The various orders correspond to the test vertices given in Eq. \eqref{eq: 16}. The CDW phases dominates most of the phase diagram. PDW phases are realized for large mass anisotropies and for a small range of angles around $\phi = \frac{n\pi}{3}$.}
    \label{fig:6VHS phase diagram}
\end{figure}

In Fig.~\ref{fig:6VHS phase diagram}, the phase with the most divergent susceptibility is shown as a function of $\eta$ and $\phi$. The various orders correspond to the test vertices from Eqs. \eqref{eq: 16}. 
As can be seen, the CDW phase dominates most of the phase diagram. The $\text{CDW}_{1,3}$ phase (in which $\langle \psi^\dag_1 \psi_3\rangle\ne 0$) dominates the area around $\eta = \frac{1}{3}$ where the particle-hole channel no longer has a logarithmic divergence but a log squared divergence on the cutoff.

The rest of the phase diagram can be understood by nesting considerations. For example, the $\text{CDW}_{4,5}$ phase dominates the area between $0<\phi<\frac{\pi}{3}$ and sufficiently large $\eta$. For small anisotropies, at the angle $\phi = \tfrac{\pi}{6}$ there is a large overlap between the energy dispersions of VH points 4 and 5, 1 and 6, and 2 and 3, which leads to a divergence in the ZS' diagram for $g_{4}$. Hence, the system will tend to develop $\rho_{4,5}, \rho_{1,6}, \rho_{2,3}$ orders. The same happens at the angle $\phi = \tfrac{\pi}{2}$ for $\rho_{1,2}, \rho_{3,4}, \rho_{5,6}$ orders, etc. 
The PDW phases are obtained either for large mass anistropy $\eta$, or small angles $\phi$ around $\frac{n\pi}{3}$ between the two copies of the VH points from the same nesting considerations. 
\subsection{Competition between orders}
The system can either develop a single $q$ component PDW (CDW) phase, a 3$q$ component PDW (CDW) or a 6$q$ component CDW. 
The system will tend toward the state that minimizes the free energy given in Eq.~\eqref{eq: 10}.
We find, just as in the case of the three VH points model, that the Ginzburg - Landau free energy is unbounded from below up to fourth order. Despite that, we can examine the free energy qualitatively to learn about the various states.
The $\text{CDW}_{1,3}$ state has six non-zero CDW components with wavevectors that connect same parity points. This case can be treated, conceptually, as two copies of the three VH points model and therefore developing all 6$q$ components minimizes the free energy.
Note that this state is generally not periodic but quasi-periodic in space. 

The $\text{CDW}_{1,2}$, $\text{CDW}_{3,6}$, $\text{CDW}_{4,5}$, $\text{PDW}_{1,2}$, $\text{PDW}_{3,6}$, and $\text{PDW}_{4,5}$ states each have three components with wavevectors that connect even and odd numbered patches. Within each state, the three components do not share the same patches, resulting in no repulsion between them ($b^C = b^P = 0$). The system will therefore develop all three components to minimize the free energy.

\section{Discussion}

\subsection{Summary of results}

We have presented two simple models of spin-polarized systems with $C_{3v}$ symmetry, and studied the effects of interactions when the system is tuned to a VH singularity.
In these two models, in the weak interaction regime, despite the diverging DOS, the system is stable if the initial interactions are repulsive.
When the initial interactions are attractive, both of the systems develop instabilities toward CDW or PDW phases, depending on the  anistropy of the dispersion at the VH point and angle between the principal axes of the dispersions of the two groups of VH points.
We note that no other orders (such as nematic order or phase separation) develop in these models, in contrast, e.g., to the Hubbard model on the square lattice near the VH singularity~\cite{Halboth2000}.
The model with three VH points develops either a single--q PDW order or 3q CDW order, depending on the anisotropy. 
close to the transition between these two phases, the system develops a 3q PDW order.
The six VH points model can develop various multiple q CDW and PDW orders depending on the mass anistropy and the angle $\phi$ between mirror symmetry related VH points.

\subsection{Comparison to other works on PDW at VH points}

Previous works studied models for multi-valley $C_3$ symmetric systems near a VH singularity that develop PDW order~\cite{Wu2023, PhysRevB.102.085103}. 
Strictly speaking, in these models, the PDW phase appears at intermediate values of the coupling, where the calculations are not fully under control.
This is since the RG flow has to take into account both terms that diverge as the log of the IR cutoff and log-squared divergent terms. 
In contrast, our model (that starts from a spin and valley polarized state) contains only log divergent singularities, and is thus under better theoretical control.
Within this model, PDW order can develop even for asymptotically weak coupling. Note, however, that the fully spin and valley polarized state itself requires a finite interaction strength; here, we have assumed that such a state exists, and studied the effects of residual interactions when a VH singularity is approached.

\subsection{Sign of the initial interaction} 

Within the patch model, an attractive initial interaction constant between the VH points is required to obtain either the CDW or the PDW phase. This leaves the question how such an attractive interaction may arise microscopically when the bare Coulombic interactions are repulsive. 
In this case, the renormalization that may lead to a change in sign in the effective inter-patch interaction has to come from degrees of freedom that are not too close to the Fermi energy. 

Conceptually, we may treat the problem in a two-step RG procedure: starting from the full electronic dispersion, we first integrate out degrees of freedom down to a small but finite energy cutoff, and then use the patch RG to renormalize the interaction further. Unfortunately, we do not know of a fully controlled way to carry out the first step of this procedure. As a crude approximation, we may use the random phase approximation (RPA) to describe the effective inter-patch interaction that serves as the initial condition for the patch RG. As we show in Appendix \ref{App E}, this indeed gives $g_1<0$ for realistic parameters for tetralayer graphene. 

\subsection{Experimental signatures}
As mentioned earlier, evidence for superconductivity in spin and valley polarized states was recently found in R4G and R5G~\cite{han2024signatures}. Moreover, a superconducting diode effect was detected in magic angle twisted trilayer graphene with zero external magnetic field, indicating possible valley polarization in the superconducting state~\cite{PDW_experimental}. 
Of course, in these materials, the interaction strength is expected to be strong, so it is not clear to what extent the weak-coupling analysis presented here is applicable. 
Nevertheless, the weak-coupling analysis may capture some of the qualitative physics. 
We conclude this section by pointing out different possible experimental signatures of different PDW states found in this work. 

The 3q PDW state breaks translational symmetry spontaneously, and is accompanied by a 3q CDW order (this follows from the $d^{\rm P}$ term in the free energy, see Eq. \eqref{eq: 10}]. Interestingly, this state supports fractional $\tfrac{h}{6e}$ quantum vortices bound to lattice dislocations~\cite{Agterberg2011,zhou2022chern}. For completeness, we present a derivation of this result in Appendix~\ref{app:fractional}. If the fractional vortices are the most stable, this state can thus be detected by magnetic scanning superconducting quantum interference device (SQUID) measurements, by comparing the number of vortices to the applied magnetic field. Even in the absence of an external magnetic field, disorder is expected to nucleate lattice dislocations that bind $\tfrac{h}{6e}$ fractional vortices.

The 1q PDW state does not break translational symmetry (since it is invariant under a combination of translation and a gauge transformation, implying that any gauge-invariant observable is also translationally invariant). Thus, in this phase, the charge density has the period of the underlying graphene lattice. However, this phase breaks $C_3$ symmetry spontaneously due to the choice of the wavevector of the PDW, implying that the critical current is anisotropic.

Finally, we note that in the weak-coupling limit treated in this work, the PDW and CDW states are expected to have gapless fermionic excitations. In particular, the PDW states have Bogoliubov Fermi surfaces~\cite{Baruch2008, berg2009striped, Brydon2018}. This is since the PDW order only gaps out states near the VH points. At intermediate to strong coupling strengths, the entire Fermi surface may become gapped.

\acknowledgements

We thank Noga Bashan, Vladimir Calvera, Andrey Chubukov, Steve Kivelson, Pavel Nosov, Oskar Vafek, and Yaar Vituri for useful discussions.
This work was supported by the European Research Council (ERC) under grant HQMAT (Grant Agreement No. 817799), by the Simons Foundation Collaboration on New Frontiers in Superconductivity (Grant SFI-MPS-NFS-00006741-03), and NSF-BSF Award No. DMR-2310312. We thank the hospitality of the Kavli Institute for Theoretical Physics, supported in part by grant NSF PHY-2309135.

\onecolumngrid
\appendix

\section{Initial value of the interaction within RPA} \label{App E}

Here, we calculate the initial value of the interaction, $g(0)$, that enters the patch RG calculation for realistic parameters for R4G, within the random phase approximation (RPA). We demonstrate that even though the bare interaction is repulsive, $g(0)$ can be negative, leading to instabilities towards CDW or PDW states. 

Our procedure can be thought of as a two-step RG process. The bare interaction is taken to be a dual-gate screened Coulomb potential, $V_{0,q} = \frac{e^2}{2\varepsilon q}\tanh(qd)$, where $d$ is the distance to the gates and $\varepsilon$ is the effective dielectric constant of the surrounding insulator. 
Integrating out high energy modes down to a small but finite cutoff $\Lambda_0$, we obtained a renormalized interaction, screened by high-energy modes. 
This renormalized interaction is then projected to the VH patches, obtaining $g(0)$. 
As a crude approximation, we use RPA to compute the effective interaction in the first step of the renormalization. 
This is justified since the original interaction is long-ranged (assuming that $d$ is larger than the inter-electron spacing in the R4G), and hence $V_{0,q}$ is strongly peaked at $q=0$. 

The RPA screened interaction is given by~\cite{PhysRevB.107.104502}:
\begin{equation}
    V_{RPA}(\vec{q}) = \frac{V_{0,q}}{1+\Pi_{0,\vec{q}}V_{0,q}},
\end{equation}
where 
$\Pi_{0,\vec{q}} = \frac{1}{A}\sum_{\vec{k}} |\Lambda_{\vec{k},\vec{q}}|^2\frac{f(\varepsilon_{\vec{k}}) - f(\varepsilon_{\vec{k}+\vec{q}})}{\varepsilon_{\vec{k}+\vec{q}} - \varepsilon_{\vec{k}}}$
is the static polarization function. $\varepsilon_{\vec{k}}$ is the energy of the electron, $f(x)$ is the Fermi-Dirac distribution and $\Lambda_{\vec{k},\vec{q}} = \braket{u_{\vec{k}}|u_{\vec{k}+\vec{q}}}$ is the overlap matrix element between states of the electron band at momenta $\vec{k}$ and $\vec{k}+\vec{q}$.
We implement a low-energy cutoff by performing the calculation at a non-zero temperature $T$. 

\begin{figure}[h]
    \centering
    \includegraphics[width= 0.9\textwidth]{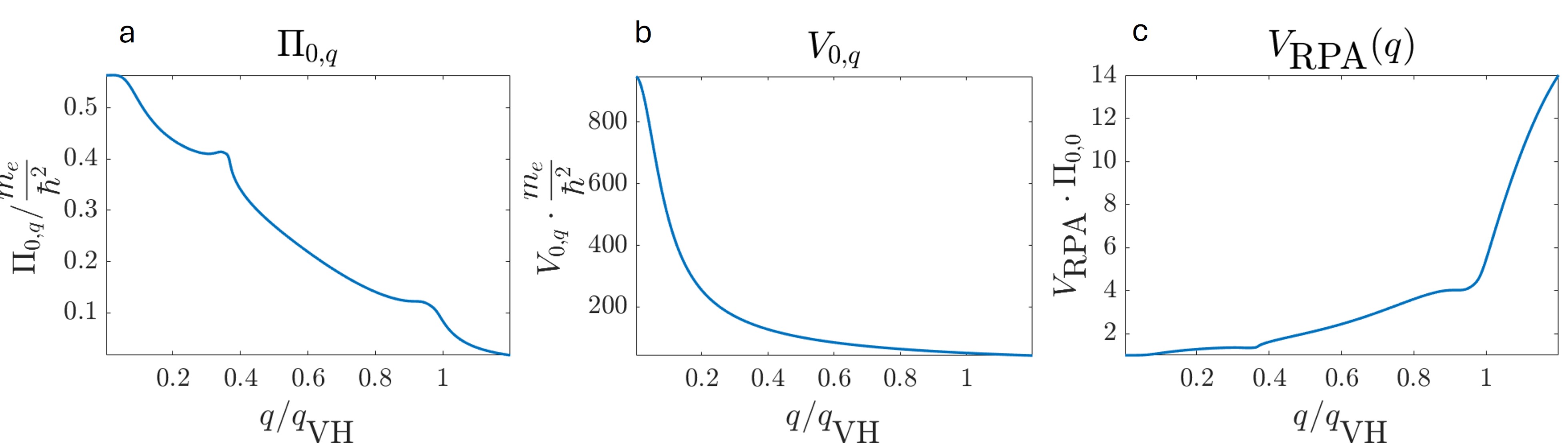} 
    \caption{\textbf{The effective interaction potential for R4G using RPA} \textbf{(a)} The static polarization calculated along the vector connecting VH points 2 and 3 (see Fig. \ref{fig: the model}) at a temperature $T=100$mK. \textbf{(b)} The screened Coulomb potential where we used $\varepsilon = 5\varepsilon_0$ and $d=40$nm. \textbf{\textbf{(c)}} The effective RPA interaction potential.}
    \label{fig: RPA potential}
\end{figure}

For the band structure of R4G, we use the model given in \cite{chou2024DasSarmaTheory}, with a displacement field correspodning to a potential difference $\Delta E =\text{70meV}$ between the top and bottom layers. The calculations were performed for the +K valley doped to the VH density at a temperature of $T = 100$mK, $\varepsilon = 5\varepsilon_0$ and $d=40$nm. We chose $\vec{q}$ as a vector pointing between VH points 2 and 3 (see Fig. \ref{fig: the model}) with a varying magnitude $q$ where $q=q_\text{VH}$ is the magnitude of vector connecting VH points 2 and 3.
Plotted in Fig. \ref{fig: RPA potential} are the static polarization function, the gate-screened Coulomb potential and the effective RPA potential. 
We find that the initial interaction constant in the three VH point model $g_1(0) = V_\text{RPA}(0) - V_\text{RPA}(q_\text{VH}) |\langle u_{\vec{q}_2} \vert u_{\vec{q}_3} \rangle|^2 < 0$ (for the R4G band structure, we find that $|\langle u_{\vec{q}_2} \vert u_{\vec{q}_3} \rangle|^2\approx 0.5$), resulting in initial attractive interactions between the VH patches.
We note that there is no divergence in the polarization at $q=0$ and at $q=q_\text{VH}$ due to the finite temperature.

\section{Three VH points RG of the interaction} \label{App A}

In this section, we will present the calculations for the RG flow equation in the three VH points model. 
The expressions for the flow coefficients defined in Eq. \eqref{eq: 5} are given by
\begin{align}
    &a_{\text{ZS'}}(\eta) = -\frac{2m\sqrt{\eta}}{\sqrt{3}\pi\left(1+\eta\right)}
    \log\left(\frac{\sqrt{3}\eta + 4\sqrt{\eta} +\sqrt{3}}{\sqrt{3}\eta - 4\sqrt{\eta} +\sqrt{3}}\right), \label{eq: A1.1}\\
    &a_{\text{BCS}}(\eta) = -\frac{4m\sqrt{\eta}}{\pi\sqrt{\left(\eta-3\right)\left(3\eta-1\right)}}
    \arctan\left(\frac{\sqrt{3}\eta - 4\sqrt{\eta} + \sqrt{3}}{\sqrt{\left(\eta-3\right)\left(3\eta-1\right)}}\right) \label{eq: A1.2},
\end{align}
where the flow parameter is $dt = d\log\left(\Lambda_\omega\right)$.
The diagrams that contribute to the beta function up to one loop order are shown in Fig.~\ref{fig:int diags}.
The contribution of the ZS diagram is given by:
\begin{equation}
    \delta^\text{ZS}g = g^2\iint_{d\Lambda} d\omega d\vec{k}\frac{1}{i\omega-\varepsilon_1(\vec{k})}\frac{1}{i\omega-\varepsilon_1(\vec{k})},
\end{equation}
where $\omega$ is the loop frequency and $\varepsilon_i(\vec{k})$ is the low energy dispersion at VH point labeled $i$. The full expression for the energy dispersion is given in Eq. \eqref{eq: 3}.  
This diagram does not contribute to the flow because both of the poles are on the same half of the plane, meaning we can always choose a contour that does not enclose them.
The ZS' contribution is given by:
\begin{equation}
    \delta^\text{ZS'}g = g^2\iint_{d\Lambda} d\omega d\vec{k}\frac{1}{i\omega-\varepsilon_1(\vec{k})}\frac{1}{i\omega-\varepsilon_3(\vec{k})}.
\end{equation}
As discussed in Chapter~\ref{Section 1B}, due to the complexity of the energy dispersion in momentum, we will renormalize the frequency:

\begin{equation} \label{eq: A.3}
    \delta^\text{ZS'}g = g^2\int_{d\Lambda_\omega} d\omega \int d\vec{k}\frac{1}{i\omega-\varepsilon_1(\vec{k})}\frac{1}{i\omega-\varepsilon_3(\vec{k})}.
\end{equation}

Transforming to polar coordinates, we denote:
\begin{equation} \label{eq: A.4}
    \varepsilon_i(\vec{k}) = \frac{k^2}{2m\sqrt{\eta}}f_i(\theta),
\end{equation}
where $f_i(\theta) = \sin^2\left(\theta + \frac{2\pi(i-2)}{3}\right) - \eta\cos^2\left(\theta + \frac{2\pi(i-2)}{3}\right)$.
Inserting this into Eq. \eqref{eq: A.3}:
\begin{align}
    \delta^\text{ZS'}g &= g^2\int_{d\Lambda_\omega} \frac{d\omega}{2\pi} \int_0^{\Lambda_k}\frac{kdk}{2\pi} \int_0^{2\pi} \frac{d\theta}{2\pi} \frac{1}{i\omega-\frac{k^2}{2m\sqrt{\eta}}f_1(\theta)}\frac{1}{i\omega-\frac{k^2}{2m\sqrt{\eta}}f_3(\theta)} = \\
    &=\frac{2m\sqrt{\eta}g^2}{2(2\pi)^3}\int_{d\Lambda_\omega}\frac{d\omega}{\omega}\int_0^{2\pi}d\theta\frac{\arctan\left(\frac{\Lambda_k^2}{2m\sqrt{\eta}\omega}f_3(\theta)\right)-\arctan\left(\frac{\Lambda_k^2}{2m\sqrt{\eta}\omega}f_1(\theta)\right)}{f_1(\theta)-f_3(\theta)} +\frac{i}{2}\frac{\log\left(\frac{\left(\frac{\Lambda_k^2}{2m\sqrt{\eta}}f_1(\theta)\right)^2+\omega^2}{\left(\frac{\Lambda_k^2} {2m\sqrt{\eta}}f_3(\theta)\right)^2+\omega^2}\right)}{f_1(\theta)-f_3(\theta)}. \notag
\end{align}
The renormalization is taken over positive and negative frequency shells eliminating the complex part of the integral. Taking the momentum cutoff to infinity, we receive:
\begin{equation}
    \delta^\text{ZS'}g = -\frac{2m\sqrt{\eta}}{16\pi^2} \int_0^{2\pi}d\theta \frac{\text{sign}(f_3(\theta))-\text{sign}(f_1(\theta))}{f_3(\theta)-f_1(\theta)} \frac{d\Lambda_\omega}{\omega}g^2.
\end{equation}
We find that the integral depends on the sign of the angular part of the dispersion. In the case of the ZS' diagram, the integral vanishes when the angular part of the both of the energy dispersion functions have the same sign. Performing the integral on opposite sign regions, we receive the result mentioned in Eq. \eqref{eq: A1.1}.

Moving to the final diagram, the contribution of the BCS diagram is given by:
\begin{equation}
    \delta^\text{BCS}g = -\frac{1}{2}g^2\iint_{d\Lambda} d\omega d\vec{k}\frac{1}{i\omega-\varepsilon_1(\vec{k})}\frac{1}{-i\omega-\varepsilon_3(\vec{k})}.
\end{equation}
Performing the calculation the same manner, we find that the BCS integral depends on the sign of the angular part of the energy dispersions. Unlike the previous case, this integral vanishes when they have opposite signs.
\begin{equation}
    \delta^\text{BCS}g = -\frac{2m\sqrt{\eta}}{32\pi^2} \int_0^{2\pi}d\theta \frac{\text{sign}(f_3(\theta))+\text{sign}(f_1(\theta))}{f_1(\theta)+f_3(\theta)} \frac{d\Lambda_\omega}{\omega}g^2.
\end{equation}
Performing the integral on equal sign regions, we receive the result mentioned in Eq. \eqref{eq: A1.2}.

\section{Three VH points RG of the test vertices} \label{App B}
\begin{figure}[h!]
  \centering
    \includegraphics[width= 0.62\textwidth]{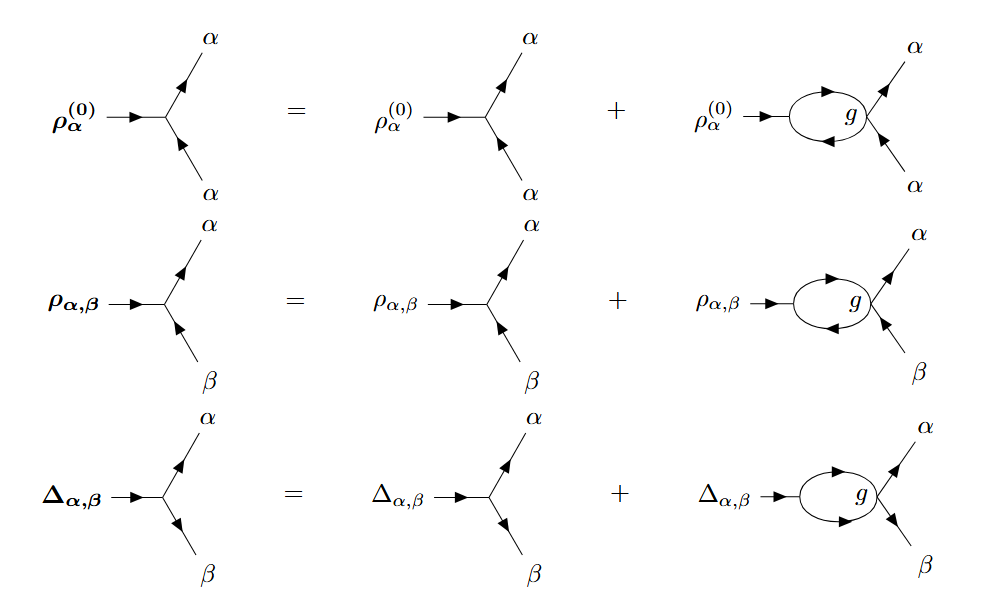} 
    \caption{The diagrams that contribute to the renormalization of the various test vertices up to one loop order. The labels on the diagrams are the labels of the VH point. The loop frequency lies in a shell of width $d\Lambda_\omega$ being integrated over.}\label{fig: sus diags}
\end{figure}

The diagrams that contribute to the RG of the test vertices are given in Fig.~\ref{fig: sus diags}.

\section{Three VH points Ginzburg Landau} \label{App C}

\begin{figure}[h]
    \centering
    \includegraphics[width= 0.7 \textwidth]{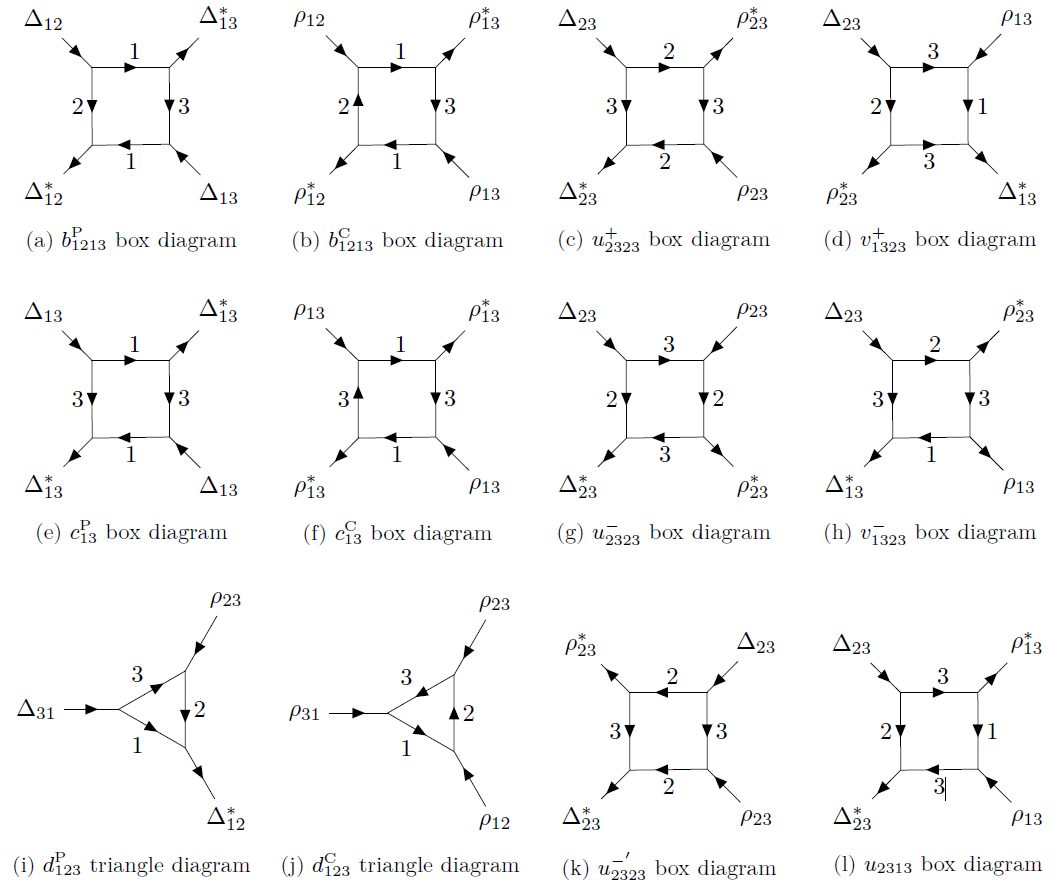} 
    \caption{Examples of the contributing diagrams to the Ginzburg-Landau free energy in the three VH points model.}
    \label{fig: GL diag}
\end{figure}

In this part, we will present the calculations of the Ginzburg-Landau free energy coefficients $b_i$, $c_i$, $d_i$, $u_{\alpha \beta, \gamma \delta}$, and $v_{\alpha \beta, \gamma \delta}$ in the three VH points model, in order to determine the leading phase under $T_c$. The contributing diagrams that we will calculate can be seen in Fig.~\ref{fig: GL diag}.
The expression for $b_\text{P}$ is given by:
\begin{equation}
    \label{eq: B.1}
    b_\text{P}(\eta) = \int_{-\infty}^{\infty} \frac{d\omega}{2\pi} \int_{-\infty}^{\infty} \frac{d^2k}{(2\pi)^2} \frac{1}{i\omega-\varepsilon_1(\vec{k})} \frac{1}{i\omega-\varepsilon_1(\vec{k})} \frac{1}{-i\omega-\varepsilon_2(\vec{k})} \frac{1}{-i\omega-\varepsilon_3(\vec{k})},
\end{equation}
where $\omega$ is the loop frequency and $\varepsilon_i(\vec{k})$ is the low energy dispersion at the VH point labeled $i$. The full expression for the energy dispersion is given in Eq.~\eqref{eq: 3}.

Following a similar calculation to the one shown in Appendix \ref{App A}, we transform to polar coordinates,and change the variable $k^2=2m\sqrt{\eta}|\omega|x$:
\begin{equation}
    b_\text{P}(\eta) = \frac{m\sqrt{\eta}}{(2\pi)^3} \int_{-\infty}^{\infty} \frac{d\omega}{|\omega|^3} \int_0^{2\pi} d\theta \int_0^\infty dx \frac{1}{i\text{sgn}(\omega) - xf_1(\theta)} \frac{1}{i\text{sgn}(\omega) - xf_1(\theta))} \frac{1}{-i\text{sgn}(\omega) - xf_2(\theta)} \frac{1}{-i\text{sgn}(\omega)- xf_3(\theta)}.
\end{equation}
Integrating over positive and negative frequencies gives the real part of the integral:
\begin{equation}
    b_\text{P}(\eta) = \frac{m\sqrt{\eta}}{(2\pi)^3} \int_{\lambda_\omega}^{\infty} \frac{d\omega}{\omega^3}  \int_0^{2\pi} d\theta \int_0^\infty dx \text{ 2Re}\left(\frac{1}{i - xf_1(\theta)} \frac{1}{i - xf_1(\theta))} \frac{1}{i + xf_2(\theta)} \frac{1}{i + xf_3(\theta)}\right),
\end{equation}
where we added a lower frequency cutoff $\lambda_\omega$. The lower cutoff $\lambda_\omega$ is equivalent to a finite temperature calculation. Performing the integral over $\omega$ and the integral over $x$ from 0 to $\Lambda_x \rightarrow\infty$, we receive:

\begin{equation}
    b_\text{P}(\eta) =\frac{m\sqrt{\eta}}{16\pi^2\lambda_\omega^2} \int_0^{2\pi} d\theta \left(\frac{f_1(f_1f_2+f_1f_3+2f_2f_3)\text{sgn}(f_1)}{(f_1+f_2)^2(f_1+f_3)^2} + \frac{f_2^2\text{sgn}(f_2)}{(f_1+f_2)^2(f_2-f_3)} + \frac{f_3^2\text{sgn}(f_3)}{(f_1+f_3)^2(f_3-f_2)}\right).
\end{equation}
The remaining angular part of this integral was calculated numerically.

The remaining Ginzburg-Landau free energy coefficients can be calculated in a similar way. These coefficients are given by: 
\begin{subequations}
    \begin{align}
      &c_\text{P}(\eta) = \frac{1}{2}\int \frac{d\omega}{2\pi}\int \frac{d^2k}{(2\pi)^2} \frac{1}{i\omega-\varepsilon_1(\vec{k})} \frac{1}{i\omega-\varepsilon_1(\vec{k})} \frac{1}{-i\omega-\varepsilon_3(\vec{k})} \frac{1}{-i\omega-\varepsilon_3(\vec{k})},\\
      &d_\text{P}(\eta) = -\int \frac{d\omega}{2\pi}\int \frac{d^2k}{(2\pi)^2} \frac{1}{i\omega-\varepsilon_1(\vec{k})} \frac{1}{i\omega-\varepsilon_2(\vec{k})} \frac{1}{-i\omega-\varepsilon_3(\vec{k})},\\
      &b_\text{C}(\eta) = \int \frac{d\omega}{2\pi}\int \frac{d^2k}{(2\pi)^2} \frac{1}{i\omega-\varepsilon_1(\vec{k})} \frac{1}{i\omega-\varepsilon_1(\vec{k})} \frac{1}{i\omega-\varepsilon_2(\vec{k})} \frac{1}{i\omega-\varepsilon_3(\vec{k})},\\
      &c_\text{C}(\eta) = \frac{1}{2}\int \frac{d\omega}{2\pi}\int \frac{d^2k}{(2\pi)^2} \frac{1}{i\omega-\varepsilon_1(\vec{k})} \frac{1}{i\omega-\varepsilon_1(\vec{k})} \frac{1}{i\omega-\varepsilon_3(\vec{k})} \frac{1}{i\omega-\varepsilon_3(\vec{k})},\\
      &d_\text{C}(\eta) = -\int \frac{d\omega}{2\pi}\int \frac{d^2k}{(2\pi)^2} \frac{1}{i\omega-\varepsilon_1(\vec{k})} \frac{1}{i\omega-\varepsilon_2(\vec{k})} \frac{1}{i\omega-\varepsilon_3(\vec{k})},\\
      &u_{2323}(\eta) = -\int \frac{d\omega}{2\pi}\int \frac{d^2k}{(2\pi)^2} \frac{1}{-i\omega-\varepsilon_3(\vec{k})} \frac{1}{-i\omega-\varepsilon_2(\vec{k})} \frac{1}{i\omega-\varepsilon_3(\vec{k})} \frac{1}{i\omega-\varepsilon_2(\vec{k})},\\
      &u_{2312}(\eta) = -\int \frac{d\omega}{2\pi}\int \frac{d^2k}{(2\pi)^2} \frac{1}{-i\omega-\varepsilon_3(\vec{k})} \frac{1}{i\omega-\varepsilon_2(\vec{k})} \frac{1}{i\omega-\varepsilon_1(\vec{k})} \frac{1}{i\omega-\varepsilon_2(\vec{k})}.      
    \end{align}
\end{subequations}
where $u_{\alpha \beta , \alpha \beta}$ are all equal to each other, $u_{\alpha \beta \neq \gamma \delta}$ are all equal to each other and $v_{\alpha \beta, \gamma \delta}=0$ due to the Gaussian averaging that cancels out.

\begin{figure}[h]
    \centering
    \includegraphics[width= 0.7 \textwidth]{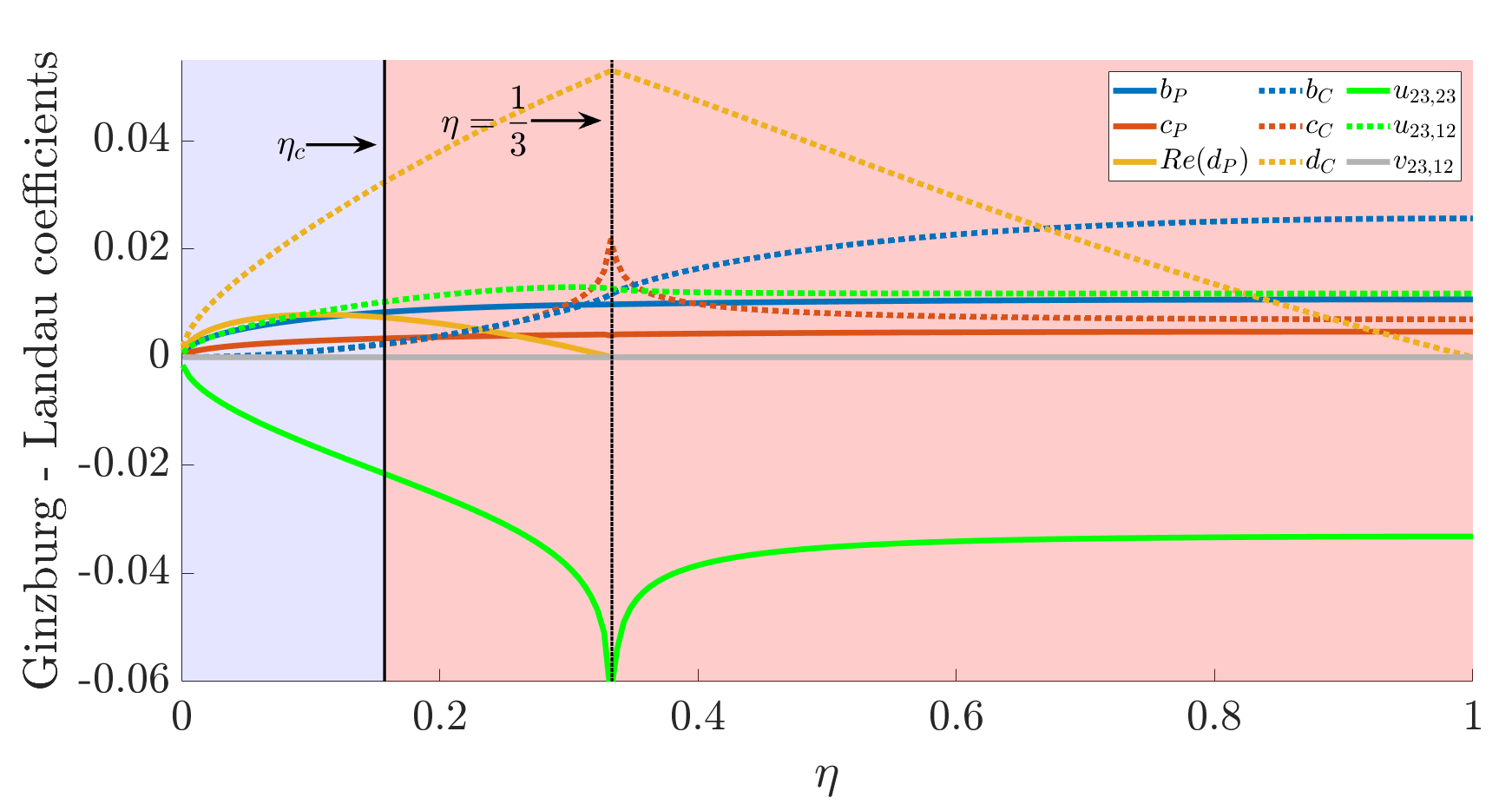} 
    \caption{The Ginzburg - Landau coefficients as a function of $\eta$ calculated from the diagrams given in Fig.~\ref{fig: GL diag}.}
    \label{fig: GL coeffs}
\end{figure}

Following the same steps as above: transforming to polar coordinates and integrating over $\omega$ and the magnitude of $\vec{k}$, we obtain:
\begin{subequations}
    \begin{align}
        &c_\text{P}(\eta) = \frac{m\sqrt{\eta}}{16\pi^2\lambda_\omega^2} \int_0^{2\pi} d\theta \left(\frac{f_1f_3\text{sgn}(f_1)}{(f_1+f_3)^3} + \frac{f_1f_3\text{sgn}(f_3)}{(f_1+f_3)^3}\right),\\
        &d_\text{P}(\eta) = -\frac{m\sqrt{\eta}}{4\pi^3\lambda_\omega} \int_0^{2\pi} d\theta \left(\frac{f_1f_2\text{log}\left(\frac{f_1}{f_2}\right)+ f_1 f_3\text{log}\left(\frac{f_1}{f_3}\right) + f_3 f_2\text{log}\left(\frac{f_3}{f_2}\right)} {(f_1-f_2)(f_1+f_3)(f_2+f_3)}\right),\\
        & b_\text{C}(\eta) =\frac{m\sqrt{\eta}}{16\pi^2\lambda_\omega^2} \int_0^{2\pi} d\theta \left(-\frac{f_1(f_1f_2+f_1f_3-2f_2f_3)\text{sgn}(f_1)}{(f_1-f_2)^2(f_1-f_3)^2} + \frac{f_2^2\text{sgn}(f_2)}{(f_1-f_2)^2(f_2-f_3)} + \frac{f_3^2\text{sgn}(f_3)}{(f_1-f_3)^2(f_3-f_2)}\right), \\
        &c_\text{C}(\eta) =\frac{m\sqrt{\eta}}{16\pi^2\lambda_\omega^2} \int_0^{2\pi} d\theta \left(- \frac{f_1f_3\text{sgn}(f_1)}{(f_1-f_3)^3} + \frac{f_1f_3\text{sgn}(f_3)}{(f_1-f_3)^3}\right),\\
        &d_\text{C}(\eta) = -\frac{m\sqrt{\eta}}{4\pi^3\lambda_\omega} \int_0^{2\pi} d\theta \left(\frac{f_1f_2\text{log}\left(\frac{f_2}{f_1}\right)+ f_1 f_3\text{log}\left(\frac{f_1}{f_3}\right) + f_3 f_2\text{log}\left(\frac{f_3}{f_2}\right)} {(f_1-f_2)(f_1-f_3)(f_2-f_3)}\right),\\
        &u_{2323}(\eta) = -\frac{m\sqrt{\eta}}{16\pi^2\lambda_\omega^2} \int_0^{2\pi} d\theta \left(\frac{f_2\text{sgn}(f_2)}{f_2^2-f_3^2} - \frac{f_3\text{sgn}(f_3)}{f_2^2 - f_3^2}\right),\\
        &u_{2312}(\eta) = \frac{m\sqrt{\eta}}{16\pi^2\lambda_\omega^2} \int_0^{2\pi} d\theta \left(\frac{f_1^2\text{sgn}(f_1)}{(f_1 - f_2)^2(f_1+f_3)} +\frac{f_2(f_2f_3-f_1f_2-2f_2f_3)\text{sgn}(f_2)}{(f_1 - f_2)^2(f_2+f_3)^2}+ \frac{f_3^2\text{sgn}(f_3)}{(f_1+f3)(f_2+f_3)^2}\right).
    \end{align}
\end{subequations}
The remaining angular part of these integrals was calculated numerically. The value of these coefficients as a function of $\eta$ is plotted in Fig.~\ref{fig: GL coeffs}.

To prove that the Ginzburg-Landau free energy is not bounded from below, we define:
\begin{equation} \label{eq: 4.10}
    F_4 = 
    \begin{pmatrix}
    |\Delta_{12}|^2\\ |\Delta_{23}|^2\\ |\Delta_{31}|^2\\ |\rho_{12}|^2\\ |\rho_{23}|^2\\ |\rho_{31}|^2
    \end{pmatrix}^{T}
    \begin{pmatrix}
    c_\text{P} & b_\text{P} & b_\text{P} & u_{1212} & u_{1223} & u_{1231}\\
    b_\text{P} & c_\text{P} & b_\text{P} & u_{2312} & u_{2323} & u_{2331}\\
    b_\text{P} & b_\text{P} & c_\text{P} & u_{3112} & u_{3123} & u_{3131}\\
    c_\text{P} & b_\text{P} & b_\text{P} & u_{1212} & u_{2311} & u_{3112}\\
    b_\text{P} & c_\text{P} & b_\text{P} & u_{1223} & u_{2323} & u_{3123}\\
    b_\text{P} & b_\text{P} & c_\text{P} & u_{1231} & u_{2331} & u_{3131}
    \end{pmatrix}_\eta
    \begin{pmatrix}
    |\Delta_{12}|^2\\ |\Delta_{23}|^2\\ |\Delta_{31}|^2\\ |\rho_{12}|^2\\ |\rho_{23}|^2\\ |\rho_{31}|^2
    \end{pmatrix}.
\end{equation}  
By looking at the eigenvalues and eigenvectors of the matrix in Eq. \eqref{eq: 4.10}  we can determine whether the quartic term in the Ginzburg - Landau free energy is unbounded from below. If there exists a negative eigenvalue, and the corresponding eigenvector has non-negative entries, the free energy is unbounded in the direction of the corresponding eigenvector. If the corresponding eigenvector includes negative entries, we need to examine its projection on each of the surfaces of the non-negative region and check whether the corresponding minor matrix has negative eigenvalues on that surface.

\section{Six VH points RG of the interaction} \label{App D}

We follow the same procedure as in Appendix~\ref{App A}, to calculate the RG flow of the interaction $g_2$. 
The expressions for the flow coefficients $A^{\text{ZS'}}(\eta,\phi)$ and $A^{\text{BCS}}(\eta,\phi)$, defined in Eqs. \eqref{eq: 15}, for $\phi<\frac{\pi}{2}$, are given by:
\begin{align}
    A^\text{ZS'}(\eta,\phi) =& \frac{m\sqrt{\eta}}{2\pi^2(\eta+1)\sin(\phi)} \log\left( \frac{\tan\left(\arcsin\left({\sqrt{\frac{\eta}{\eta+1}}+\frac{\phi}{2}}\right)\right)} {\tan\left(\arcsin\left({\sqrt{\frac{\eta}{\eta+1}}-\frac{\phi}{2}}\right)\right)}\right), \label{eq: 4.14}\\
    A_2^\text{BCS}(\eta,\phi) =& \frac{m\sqrt{\eta}}{\pi^2(\eta+1)\alpha_\text{BCS}}\text{arctanh}\left( \frac{\left(\frac{\eta-3}{\eta+1}-\cos(2\phi)\right)\tan(\arcsin\left(\sqrt{\frac{\eta}{\eta+1}}\right)-\phi) - \sin(2\phi)}{\alpha_\text{BCS}}\right)\notag\\
    &+\frac{m\sqrt{\eta}}{\pi^2(\eta+1)\alpha_\text{BCS}}\text{arctanh}\left( \frac{\left(\frac{\eta-3}{\eta+1}-\cos(2\phi)\right)\sqrt{\eta} + \sin(2\phi)}{\alpha_\text{BCS}}\right), \label{eq: 4.15} 
\end{align}
where $\alpha_\text{BCS}=\sqrt{2\left(\cos(2\phi)-2\left(\frac{2\eta}{\eta+1}\right)^2+\frac{8\eta}{\eta+1}-1\right)}$.

\section{Fractional vortices in 3q PDW states}
\label{app:fractional}
An analysis of the fundamental defects of a three-component PDW appeared
in Ref.~\cite{Agterberg2011}. We provide a simple derivation of the existence of
$\tfrac{h}{6e}$ vortices for completeness.

\begin{figure}[h]
\includegraphics[width=0.3\columnwidth]{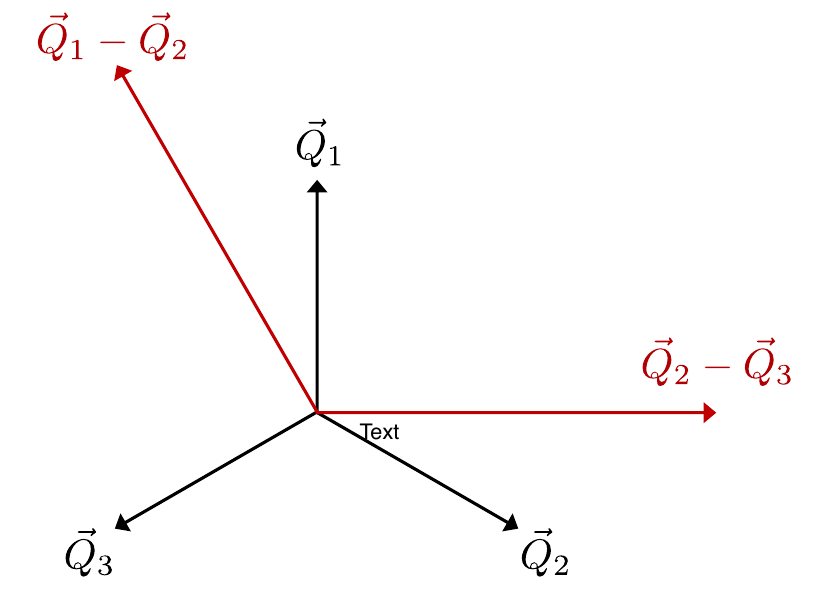}\caption{The vectors $\vec{Q}_{1,2,3}$ of the three components of the PDW.
The two vectors $\vec{Q}_{1}-\vec{Q}_{2}$ and $\vec{Q}_{2}-\vec{Q}_{3}$
are reciprocal lattice vectors of the lattice corresponding to the
modulation of the charge in the PDW state.}
\label{fig:Q}
\end{figure}

The PDW order parameter can be written in real space as
\begin{equation}
\Delta(\vec{x})=\left|\Delta\right|\sum_{i=1}^{3}e^{i\vec{Q}_{i}\cdot\vec{x}+i\varphi_{i}},
\end{equation}
where the three PDW components have the same magnitude, $|\Delta|$,
by $C_{3v}$ symmetry. The wavevectors $\vec{Q}_{1,2,3}$ are given
in Fig.~\ref{fig:Q}, and are related to the wavevectors of the VH locations,
$\vec{q}_{1,2,3}$, by: $\vec{Q}_{1}=\vec{q}_{2}+\vec{q}_{3}$, $\vec{Q}_{2}=\vec{q}_{1}+\vec{q}_{3}$,
$\vec{Q}_{3}=\vec{q}_{1}+\vec{q}_{2}$. Table \ref{tab:sym} lists
how the phase fields $\varphi_{i}$ transform under the symmetries
of the system. 

\begin{table}[h]
  \caption{Transformation properties of the phase fields under symmetry operations.}
  \label{tab:sym}

  \centering
\begin{tabular}{|c|c|c|}
\hline 
Charge $U(1)$ & Translation by $\vec{a}$ & $C_{3}$\tabularnewline
\hline 
$\varphi_{i}\rightarrow\varphi_{i}+2\alpha,$ & $\varphi_{i}\rightarrow\varphi_{i}+\vec{Q}_{i}\cdot\vec{a}$ & $\varphi_{1}\rightarrow\varphi_{2}$\tabularnewline
$\vec{A}\rightarrow\vec{A}-\frac{\hbar}{e}\nabla\alpha.$ &  & $\varphi_{2}\rightarrow\varphi_{3}$\tabularnewline
 &  & $\varphi_{3}\rightarrow\varphi_{1}$\tabularnewline
\hline 
\end{tabular}

\end{table}

In order to write the free energy, it is convenient to change variables to 
\begin{align}
\phi_{0} & =\varphi_{1}+\varphi_{2}+\varphi_{3},\nonumber \\
\phi_{1} & =\varphi_{1}-\frac{1}{2}\left(\varphi_{2}+\varphi_{3}\right),\nonumber \\
\phi_{2} & =\varphi_{2}-\varphi_{3}.\label{eq:phi}
\end{align}
Here, $\phi_{0}$ is the phase of the condensate which transforms
under a charge $U(1)$ transformation, and $\phi_{1,2}$ are proportional
to the two components of the displacement vector $\vec{u}=(u_{1},u_{2})$
along two orthogonal directions. Note that $\phi_{0}$ is translationally
invariant, whereas $\phi_{1,2}$ are charge neutral, i.e., they do
not transform under charge $U(1).$ In terms of these variables, the
free energy can be written as
\begin{equation}
F=\int d^{2}x\,\frac{1}{2}\kappa_{P}\left(\vec{\nabla}\phi_{0}+\frac{6e}{\hbar}\vec{A}\right)^{2}+F_{el}[\phi_{1},\phi_{2}].\label{eq:F}
\end{equation}
Here, $\kappa_{P}$ is the phase stiffness, and $F_{el}$ is the elastic
energy of the PDW:
\begin{equation}
F_{el}[\phi_{1},\phi_{2}]=\int d^{2}x\,\Big(\frac{\kappa_{0}}{2}\left[\partial_{1}\phi_{1}+\partial_{2}\phi_{2}\right]^{2}+\frac{\kappa_{1}}{2}\left[\left(\partial_{1}\phi_{1}-\partial_{2}\phi_{2}\right)^{2}+\left(\partial_{1}\phi_{2}+\partial_{2}\phi_{1}\right)^{2}\right]\Big),
\end{equation}
where $\kappa_{0}$ and $\kappa_{1}$ are elastic constants related
to the bulk and shear moduli of the PDW crystal. The spatial derivatives
$\partial_{1,2}$ are taken in the direction of the displacements
corresponding to $\phi_{1,2}$. 

Requiring that the order parameter is single-valued, the three phases
have to satisfy \begin{equation}
\oint\vec{\nabla}\varphi_{1}\cdot d\vec{x}=2\pi n_{i}
\end{equation}
with integer $n_{i}$ for any closed curve. Consider an elementary
point defect around which $(n_{1},n_{2},n_{3})=(1,0,0)$. Using Eq.
(\ref{eq:phi}), we find that for a curve around such a defect, 
\begin{align}
\oint\vec{\nabla}\phi_{0}\cdot d\vec{x} & =2\pi,\nonumber \\
\oint\vec{\nabla}\phi_{1}\cdot d\vec{x} & =2\pi,\nonumber \\
\oint\vec{\nabla}\phi_{2}\cdot d\vec{x} & =0.
\end{align}
From (\ref{eq:F}), we see that this corresponds to a flux $\Phi=\oint \vec{A}\cdot d\vec{x}=-\frac{\hbar}{6e}\oint\vec{\nabla}\phi_{0}\cdot d\vec{x}=-\frac{h}{6e}$.
Therefore, this defect corresponds to a flux $|\Phi|=\frac{h}{6e}$
bound to an elementary lattice dislocation with Burgers vector $(a_{c},0)$,
where $a_{c}$ is the triangular lattice corresponding to the charge
modulation in the PDW state (or, equivalently, to the periodicity
of $|\Delta(\vec{x})|^{2}$). 

\twocolumngrid

\bibliography{refs_fixed}

\end{document}